\documentclass[11pt,a4paper]{article}
\usepackage[sort]{cite}
\usepackage{url,colortbl}
\usepackage{xcolor}
\usepackage{bm,amsmath,amssymb}
\usepackage[]{graphicx}
\usepackage[normalem]{ulem}
\usepackage{bbding}
\usepackage{enumitem}
\usepackage{duckuments}
\usepackage{subfig}
\usepackage[margin=2.5cm]{geometry}
\graphicspath{{./Images/}}
\usepackage{authblk}
\usepackage{xr}
\usepackage[pdfencoding=auto, psdextra]{hyperref}
\usepackage{amsmath,amssymb,amsfonts}
\usepackage{bm}
\usepackage{amsmath,amssymb,amsfonts,amsthm}
\usepackage{bm}
\usepackage{mathtools}
\usepackage{geometry}
\usepackage{graphicx}
\usepackage{hyperref}
\usepackage{algorithm}
\usepackage{algpseudocode}
\usepackage{booktabs}
\usepackage{mathrsfs}
\usepackage{etoolbox}    

\usepackage{siunitx}     
          
\usepackage{physics}     

\newcommand{\vhat}{\hat{\mathbf v}}

\begin{document}

\title{Inverse statistics of active matter trajectories to distinguish interaction kernel anisotropy from emergent correlations}
\author[1]{Simon F. Martina-Perez\thanks{Email: simon.martina-perez@medschool.ox.ac.uk}}
\affil[1]{School of Medicine and Biomedical Sciences, University of Oxford, UK}

\date{}
\maketitle

\begin{abstract}
High‑resolution imaging provides dense trajectories of migrating cells, flocking animals, and synthetic active particles, from which interaction laws can be determined with a wide variety of methods. Yet, distinguishing whether front-back or lateral biases seen in such data reflect intrinsic anisotropy in the interaction kernel or emergent correlations that are nevertheless produced by isotropic pairwise interaction forces remains an open challenge. We resolve this ambiguity by deriving a linear partial differential equation that connects measurable two‑point velocity correlations to an unknown, distance‑ and angle‑dependent interaction kernel. Turing-like instabilities can occur which allows for dipolar or quadrupolar patterns to arise even when agents interact according to an underlying attraction-repulsion law that is angle-independent. We then show that incorporating a weak velocity-alignment force can interfere with anisotropic pattern formation by suppressing dipolar patterns. We validate these predictions with agent-based simulations and provide design guidance for experiments that seek to discriminate intrinsic anisotropy from emergent effects. 
\end{abstract}

\maketitle
\section{Introduction}
Collective motion of individual agents in living and artificial systems is often captured by models in which self-propelled particles (SPP) interact through pairwise forces that depend only on their mutual distance \cite{vicsek1995novel,couzin2002collective,bertozzi2006self,szabo2006phase}. Despite their simplicity, canonical SPP models recapitulate a wide repertoire of emergent patterns that are ubiquitous in nature: flocking, milling, phase-separated clustering and emergent ordering~\cite{vicsek1995novel,bertozzi2006self}.  All of these structures have clear biological analogues: rotating mills of fish, coherent flocks of birds, and the collective streaming observed in epithelial sheets during wound healing \cite{couzin2002collective,szabo2006phase}. The phenomenological success of these models has long motivated the
inverse problem: inferring the underlying interaction laws directly from experimentally observed trajectory data with the aim of uncovering their mechanistic basis. One question of particular interest is determining whether agents exchange information in a direction-dependent (anisotropic) manner~\cite{HUTH1992365,couzin2005effective,herbert2011inferring,heras19attention,rosenthal2015revealing}.

A common strategy to detect anisotropy in the rules governing interactions between agents is to compute two‐point spatial or velocity correlations and test whether the field is isotropic.  Bayesian model comparison of such statistics has been used to quantify directional bias with minimal modelling assumptions~\cite{GiuggioliLuca2015DRaB,JiangLi2017Iini}. Further, mapping the locations of cell-cell contacts on the cell membrane and mapping subsequent velocity changes has shown how electrotactic stimuli disrupt the spatial patterns of cell-cell interactions among human corneal epithelial cells \cite{crossley2025electrotaxis}. Correlation methods are appealing for high‐throughput imaging assays, but they do not by themselves reveal the underlying forces driving any observed anisotropy. Over the last decade a range of statistical and machine-learning tools have been developed for this task. Rather than review all such methods, we highlight a few representative methods. Sparse identification of nonlinear dynamics (SINDy) \cite{brunton2016discovering} and its weak-form variant WSINDy have been extended to second-order agent systems; a recent study used WSINDy to learn anisotropic interaction laws in heterogeneous cell monolayers and automatically clustered cells into behavioural subtypes~\cite{messenger2022learning}. Non-parametric regression frameworks couple consistency theorems with efficient algorithms to recover distance-based kernels from many trajectories~\cite{lu2019pnas,lu2021jmlr}. Together these approaches provide interpretable force laws, but they often incorporate angular basis functions by default, implicitly assuming that any directional bias visible in the data reflects a directional force. Moreover, deep learning based attention network approaches have recently been applied to infer directional interaction rules in both animal and cellular systems: for example, deep attention networks uncovered the anisotropic social forces guiding zebrafish shoals \cite{heras19}, and analogous architectures have been used to learn direction‐dependent cues in collective cell migration \cite{lachance22}. Similar deep attention frameworks have also facilitated the identification of invasion and migration genes in neural crest and melanoma cells from high‐throughput screens \cite{kulesa2025devdyn}.

This assumption is problematic, because, as will be shown in this work, anisotropic correlations can emerge even when the true interaction kernel is strictly isotropic. Sampling artefacts and uneven neighbour distributions can induce spurious angular structure even when the true kernel is isotropic. Spurious detection of anisotropic patterns in truly isotropic force fields is not confined to kinetically interacting agents. For instance, in seismic surface-wave tomography (a geological application) uneven ray coverage around sharp velocity contrasts produces spurious angular anisotropy, despite the Earth’s material being isotropic \cite{PaulAnne2024Mais}. Incorporating these artefacts as genuine directional forces not only over‐complicates the model but also masks the true driving mechanisms.  Hence, anisotropic interactions should be introduced only when the data provide clear evidence for them. This work uses a kinetic theory-based analysis to determine under what conditions two‐point velocity correlations reliably reflect intrinsic directional biases in an otherwise isotropic SPP framework.  The main findings are that observing anisotropic statistics does not, by itself, justify baking anisotropy into the governing equations.  Instead, one must first rule out alternative origins of apparent directionality, such as persistence, inertial delays, or nonlinear density pattern formation, before attributing it to anisotropic forces.  

The structure of this work is as follows. Starting from a canonical SPP model, we derive a closed Fokker-Planck equation for the two-particle velocity probability density is derived, and a Kramers-Moyal expansion \cite{Risken1996,Kramers1940,Moyal1949} is used to find the governing equation for the spatial correlation field in Section~\ref{sec:forward}. Linear stability analysis of the resulting partial differential equation (PDE) in Section~\ref{sec:emergence} shows that front-back or lateral biases can emerge in this correlation even though the interaction kernel is angle-independent, and that dipolar interaction kernels can suppress a dipolar instability in the emergent correlation pattern. Finally, in Section~\ref{sec:alignment}, in addition to isotropic attraction-repulsion kernels, we also incorporate an alignment term and show that it can selectively shift the dispersion relations of some of the unstable correlation modes, allowing us to probe how alignment suppresses or enhances angular instabilities.

\section{Governing equations of velocity correlations}\label{sec:forward}
The purpose of this section is to derive a closed-form PDE for the velocity-velocity correlation field of two interacting particles. In Section~\ref{sec:Fokker-Planck} the Fokker-Planck equation describing the probability density for the velocities of two interacting particles is derived. Then, in Section~\ref{sec:velocity-correlation} a closed-form expression for a correlation function in the frame of motion of a single particle is found by using the Fokker-Planck equation. This is used in Sections~\ref{sec:first-order-modes}-\ref{sec:minimal_anisotropy}, which investigate the conditions under which anisotropy can arise in the correlation function, and show that this can occur with a simple, isotropic interaction kernel. 

\subsection{Pair‑density formulation}
\label{sec:Fokker-Planck}
This work considers the interactions between $N$ identical particles with positions $\mathbf x_i(t)$ and velocities $\mathbf v_i(t)$ that obey the following dynamics,
\begin{align}
  \dot{\mathbf x}_i &= \mathbf v_i, \label{eq:kinematics}\\[2pt]
  \dot{\mathbf v}_i &= S(\mathbf v_i)\,\mathbf v_i
  + \frac{1}{N}\sum_{j\neq i} f_{ar}(r_{ij}, \phi_{ij})\,\hat{\mathbf r}_{ij}
  + \sigma\,\boldsymbol\xi_i(t). \label{eq:dynamics}
\end{align}
Here, $r_{ij}= \lVert\mathbf x_j-\mathbf x_i\rVert$, 
$\hat{\mathbf r}_{ij}= (\mathbf x_j-\mathbf x_i)/r_{ij}$, and and $\phi_{ij}$ is the angle between $\mathbf{v}_i$ and $\mathbf{x}_j - \mathbf{x}_i$.
The scalar self-propulsion term $S(\mathbf v_i)$ drives each
particle toward a preferred speed and can be chosen to recover, for
example, the persistent-random-walk of migrating cells
\cite{szabo2006phase}. The model allows the pairwise attraction-repulsion force, $f_{ar}(r_{ij},\phi_{ij})$, to depend on the separation vector, $\phi_{ij}$, and thus be anisotropic. Nevertheless, a commonly made assumption is that of an isotropic interaction force, reflecting the fact that many cells or animals sense neighbours primarily through distance cues such as adhesive contacts, chemotactic gradients or visual range.  Finally,
$\boldsymbol\xi_i$ is standard Gaussian white noise with amplitude
$\sigma$. The pair probability density, $P$, is given by
\begin{equation}\label{eq:pair_pdf}
P(\mathbf r,\mathbf v,\mathbf v',t)
  = \bigl\langle
      \sum_{i\neq j}
      \delta\bigl(\mathbf r-\mathbf x_j+\mathbf x_i\bigr)\,
      \delta\bigl(\mathbf v-\mathbf v_i\bigr)\,
      \delta\bigl(\mathbf v'-\mathbf v_j\bigr)
    \bigr\rangle .
\end{equation}
Standard BBGKY reduction combined with a mean‑field closure (also called molecular-chaos closure) shows that the pair probability density, $P$, obeys the Fokker-Planck equation ~\cite{cercignani_book}, which is given by
\begin{equation}\label{eq:Fokker-Planck_pair}
\partial_t P
  + (\mathbf v'-\mathbf v)\!\cdot\!\nabla_{\mathbf r} P
  + (\mathcal{L}_{\mathbf v}+\mathcal{L}_{\mathbf v'})[P]
  = \frac{\sigma^2}{2}\bigl(
      \Delta_{\mathbf v} + \Delta_{\mathbf v'}
    \bigr) P ,
\end{equation}
where the operator $\mathcal{L}_{\mathbf v}[P]$ groups together the deterministic self‑propulsion drift and the mean‑field attraction-repulsion term.  See Appendix~\ref{app:pair_fp_full} for the derivation of the Fokker-Planck equation and more details.

\subsection{Velocity-velocity correlation field}
\label{sec:velocity-correlation}
Given that particles move at near-constant velocity owing to the forcing term, $S$, the pair density, $P$, can be projected onto a fixed speed, $v=v'=v_0$, thus eliminating the fast radial dynamics of the speeds and focusing solely on the slower angular coordinates:
\begin{equation}
\label{eq:shell_projection}
  \tilde P(r,\varphi,\varphi',t)
  := \int_0^\infty\!\!\int_0^\infty
       P(r,\mathbf v,\mathbf v',t)\,
       \delta(v-v_0)\delta(v'-v_0)\,v\,v'\,dv\,dv'.
\end{equation}
All angular integrations below refer to $\tilde P$. In the co‑moving polar frame of a focal particle, $i$, with $\theta=0$ directed along its instantaneous velocity, the equal‑time velocity-velocity correlation, $C$, is defined by
\begin{equation}\label{eq:C_def}
C(r,\theta,t)
   = \Bigl\langle
       \vhat_i(t)\!\cdot\!\vhat_j(t)\;
     \Bigr|\;
     r_{ij}=r,\;\phi_{ij}=\theta
     \Bigr\rangle.
\end{equation}
Multiplying the Fokker-Planck equation \eqref{eq:Fokker-Planck_pair} by
$\vhat\!\cdot\!\vhat'$ and integrating over angular variables yields,
\begin{align}\label{eq:C_evolution}
\partial_t C
&= -v_0\,\partial_r C
   + D_r\bigl(\partial_r^2 + r^{-1}\partial_r\bigr)C
   + \frac{D_\theta}{r^2}\partial_\theta^2 C
   - 2\gamma\,C
   + S_{\mathrm{ar}}(r,\theta), \\[2pt]
S_{\mathrm{ar}}(r,\theta)
&= \rho\,
   g(r,\theta)\,
   f_{\mathrm{ar}}(r,\theta),
\end{align}
where \(g(r,\theta)\) is the neighbour distribution around a focal particle, and \(\rho\) is the global number density. The coefficients \(v_0,\,\gamma=\sigma^2/2v_0^2,\,D_r,\,D_\theta\) are single‑particle kinetic statistics, representing, respectively: the mean self-propulsion speed; the angular decorrelation rate; the radial diffusivity of the pair separation; and the angular diffusivity of the separation vector. For clarity of exposition, we omit the derivation of Equation~\eqref{eq:C_evolution} here, see Appendix~\ref{app:cond_corr_full} for the derivation and further details.

\section{Emergence of correlated dipoles and quadrupoles}
\label{sec:emergence}
The linear dynamics of the correlation field, $C$, are governed by fluctuations of the pair density projected at constant speed, which, when writing $\theta=\arg\mathbf v$ and $\theta'=\arg\mathbf v'$ for headings,
and $\varphi=\arg\mathbf r$ for the polar angle of the separation vector, is given by
\begin{equation}
  \tilde P(r,\theta,\theta',\varphi,t)
  :=\!\int_{0}^{\infty}\!\!\int_{0}^{\infty}
        P\!\bigl(r,\mathbf v,\mathbf v',t\bigr)\,
        \delta(v-v_{0})\,\delta(v'-v_{0})\,v\,v'\,dv\,dv' .
\end{equation}
At this point, it is important to notice that inspection of
\begin{equation}
  C(r,\Theta,t)=
  \frac{\displaystyle\iint
          \cos(\theta-\theta')\,
          \tilde P(r,\theta,\theta',\theta+\Theta,t)\,d\theta\,d\theta'}
       {\displaystyle\iint
          \tilde P(r,\theta,\theta',\theta+\Theta,t)\,d\theta\,d\theta'},
\end{equation}
shows that any non‑isotropic perturbation of $\tilde P$ feeds directly into the rise of angular structure in $C$.  Linear stability is therefore most naturally performed on $\tilde P$ itself.

\subsection{Linear stability of pair correlations}
\label{sec:first-order-modes}
Projection of the Fokker-Planck pair equation onto
$v=v'=v_{0}$, together with
$\hat{\mathbf v}=(\cos\theta,\sin\theta)$ and
$\hat{\mathbf v}'=(\cos\theta',\sin\theta')$, yields the governing equation for the projected pair distribution, $\tilde P$,
\begin{equation}
  \partial_t \tilde P
  + \frac{v_0}{r}\Bigl[
        \sin(\theta'-\varphi)\,\partial_\theta\tilde P
      - \sin(\theta-\varphi)\,\partial_{\theta'}\tilde P
    \Bigr]
  - v_0\cos(\theta'-\varphi)\,\partial_r\tilde P
  = \widehat{\mathcal L}_r\tilde P
     + D_\theta\bigl(
         \partial_{\theta\theta}
       + \partial_{\theta'\theta'}
       \bigr)\tilde P,
  \label{eq:fixed-speed}
\end{equation}
where $D_\theta = \frac{\sigma^{2}}{2}$, $\widehat{\mathcal L}_r\tilde P
  := -\partial_r\!\bigl[F(r)\,\tilde P\bigr]$, and $F(r)$ is an isotropic attraction-repulsion kernel. For a spatially homogeneous suspension, the stationary solution is
\[
  \tilde P_{0}(r,\theta,\theta')=\frac{\rho_{0}}{(2\pi)^{2}},
\]
where $\rho_{0}$ is the bulk number density. Here, $\tilde P_{0}$ is independent of all angular and radial variables and normalised so that
\(\int \tilde P_{0}\,d\theta\,d\theta'=1\). The subsequent linear stability analysis will investigate the growth of angular modes in \(\tilde P\). The existence of instabilities of such Fourier blocks maps one‑to‑one onto the emergence of front-back patterns in \(C\). Write \(\tilde P=\tilde P_{0}+\varepsilon\,\delta\tilde P\) and expand
the perturbation in angular Fourier modes and the angularly averaged radial basis, \textit{i.e.}, take the Hankel transform,
\begin{equation*}
    \delta\tilde P(r,\theta,\theta’,t)
=\sum_{m,n\in\mathbb Z}\int_0^\infty k\,\widehat P_{mn}(k,t)\,J_0(kr)\,e^{im\theta}e^{in\theta’}\,\dd k,
\end{equation*}
Note that the Hankel transform is equivalent to averaging the plane wave $e^{ik r\cos\varphi}$ over its dominant direction to obtain $J_0$. Since $P_0$ is constant, perturbations obey
\begin{equation}
\label{eq:perturbations_governing}
\partial_t \delta\tilde P
= \mathcal L_r\,\delta\tilde P
+ D_\theta(\partial_{\theta\theta}+\partial_{\theta'\theta'})\delta\tilde P
- v_0\cos(\theta'-\varphi)\,\partial_r\delta\tilde P
+\frac{v_0}{r}\big[\sin(\theta'-\varphi)\,\partial_\theta-\sin(\theta-\varphi)\,\partial_{\theta'}\big]\delta\tilde P.
\end{equation}
Multiplying Equation~\eqref{eq:perturbations_governing} by $e^{-im\theta}e^{-in\theta'}$ and integrating over $\theta, \theta'$ allows to produce the following evolution equations. First, using that $\sin(\bullet-\varphi)=(1/2i)\big(e^{i(\bullet-\varphi)}-e^{-i(\bullet-\varphi)}\big)$ and $\partial_\bullet e^{im\bullet}=im\,e^{im\bullet}$ one obtains that upon projecting Equation~\eqref{eq:perturbations_governing} the following terms map onto nearest-neighbour couplings between angular modes,
\begin{align*}
    \frac{v_0}{r}\sin(\theta’-\varphi)\,\partial_\theta \;&\leadsto\;
\frac{v_0}{2r}\,m\big(\widehat P_{m,n-1}e^{+i\varphi}-\widehat P_{m,n+1}e^{-i\varphi}\big),\\
-\frac{v_0}{r}\sin(\theta-\varphi)\,\partial_{\theta’} \;&\leadsto\;
-\frac{v_0}{2r}\,n\big(\widehat P_{m-1,n}e^{+i\varphi}-\widehat P_{m+1,n}e^{-i\varphi}\big).
\end{align*}
Observing that
\(
  \cos(\theta-\varphi)e^{in\theta'}
   =\tfrac12\bigl(e^{i(n+1)\theta'}+e^{i(n-1)\theta'}\bigr)e^{-i\varphi}
\),
the cosine term in Equation~\eqref{eq:perturbations_governing}, upon projection, maps as 
\[
v_0\cos(\theta’-\varphi)\,\partial_r \delta\tilde P
\ \leadsto\
\frac{v_0}{2}\Big[
e^{+i\varphi}\, \partial_r\!\big(k\,\widehat P_{m,n-1}(k,t)\,J_0(kr)\big)
+
e^{-i\varphi}\, \partial_r\!\big(k\,\widehat P_{m,n+1}(k,t)\,J_0(kr)\big)
\Big].\]
The radial operator diagonalises such that
\(\widehat{\mathcal L}_r\!\to\!\lambda_{r}(k)\) with
\[
  \lambda_{r}(k)= -k^{2}D_{t} + \rho_{0}k^{2}\widehat F(k),\qquad
  D_{t}:=\frac{\sigma^{2}}{2},\qquad
  \widehat F(k)=2\pi\int_{0}^{\infty}rF(r)J_{0}(kr)\,dr .
\]
Angular diffusion contributes $-D_\theta(m^2+n^2)$. Collecting terms, projection of Equation~\eqref{eq:perturbations_governing} produces, for each $(m,n)$,
\begin{align}
\partial_t\big(k\widehat P_{mn}J_0\big)
&= kJ_0\Big(\lambda_r(k)-D_\theta(m^2+n^2)\Big)\widehat P_{mn} \\
&\quad + \frac{v_0}{2r}\,m\Big(k\widehat P_{m,n-1}J_0\,e^{+i\varphi}-k\widehat P_{m,n+1}J_0\,e^{-i\varphi}\Big)\\
&\quad - \frac{v_0}{2r}\,n\Big(k\widehat P_{m-1,n}J_0\,e^{+i\varphi}-k\widehat P_{m+1,n}J_0\,e^{-i\varphi}\Big)\\
&\quad + \frac{v_0}{2}\Big[
e^{+i\varphi}\,\partial_r\!\big(k\widehat P_{m,n-1}J_0\big)
+e^{-i\varphi}\,\partial_r\!\big(k\widehat P_{m,n+1}J_0\big)
\Big].
\end{align}
To eliminate dependence on $\varphi$, define the next projection 
\[\langle f\rangle_\varphi
:=\frac{1}{2\pi J_0(kr)}\int_{0}^{2\pi}
f(\varphi)\,e^{ik r\cos\varphi}\,d\varphi .\]
Noting that \(\langle e^{\pm i\varphi}\rangle_\varphi=\pm i\,J_1(kr)/J_0(kr)\),  \(\langle 1\rangle_\varphi=1\), and \(J_0’(x)=-J_1(x)\). Collecting terms, and using the Ansatz $\widehat P_{mn}(k,t)=\widehat A_{mn}(k)\,e^{\lambda t}$, for each $(m,n)$ the evolution is given by
\begin{equation}
  \bigl[\lambda-\lambda_{r}(k)+D_\theta(m^{2}+n^{2})\bigr]\,\hat P_{mn}
  +\beta\!\bigl(
       \hat P_{m-1,n}-\hat P_{m+1,n}
      -\hat P_{m,n-1}+\hat P_{m,n+1}
     \bigr)=0.
\end{equation}
Here, we have used the approximation that $kr \approx 1$ so that
$v_0 k J_1(kr)/J_0(kr) \approx v_0 k^2 r/2$. The constant $\beta$ is therefore given by $\beta = k v_0/2$.

\subsubsection{The isotropic block, \((m,n)=(0,0)\)}
Since $\hat{P}_{-1,0} = \hat{P}_{1,0}^{*}$ and $\hat{P}_{0,-1} = \hat{P}_{0,1}^{*}$, the isotropic mode decouples, so that
\[
  \lambda_{0}(k)=\lambda_{r}(k)
  = -k^{2}D_{t}+\rho_{0}k^{2}\widehat F(k).
\]
A clustering instability appears when \(\lambda_{0}(k)>0\), \textit{i.e.},
\begin{equation}
  \rho_{0}\widehat F(k)\;>\;D_{t}=\frac{\sigma^{2}}{2}.
  \label{eq:isotropic-instability}
\end{equation}
In this case, an instability arises through the eigenvalue $\lambda_0(k)$ as it balances translational diffusion against the Fourier transform of the attraction-repulsion kernel. If the condition in Equation~\eqref{eq:isotropic-instability} occurs, then density modulations at that wavenumber grow exponentially. In that case, particles aggregate into clusters that are sill isotropic in angle. It is only after this clustering sets in that polar patterns can arise due to contributions from $\beta$.

\subsubsection{The first anisotropic instability, \((m,n)=\pm1,0\)}
The $4\times4$ block spanned by
$\{(1,0),(-1,0),(0,1),(0,-1)\}$ carries the minimal information needed to distinguish front from back in anisotropic patterns, therefore it will be referred to as the first anisotropic instability.  The dominant eigenvalue is given by
\[
  \lambda_{\pm1}(k)
  = \lambda_{0}(k)
    - D_\theta
    + \frac{\beta^{2}}{\lambda_{0}(k)+D_\theta},
\]
with
$\beta=k v_{0}/2$ as before, and $D_\theta=\sigma^{2}/2$. Positivity of $\lambda_{\pm1}$ is equivalent to the condition that
\(\lambda_{0}(k)>D_\theta\) as well as \(\beta^{2}>D_\theta\bigl[\lambda_{0}(k)-D_\theta\bigr]\). Now, substitution of $\lambda_{0}$ and $\beta$ produces,
\begin{align}
  \rho_{0}\widehat F(k)
   &> \frac{\sigma^{2}}{2}\!
      \Bigl(1+\frac{1}{k^{2}}\Bigr),\\[2pt]
  k^{2}v_{0}^{2}
   &> \frac{\sigma^{2}}{2}\!
      \Bigl[2\rho_{0}\widehat F(k)-\sigma^{2}
            -\frac{\sigma^{2}}{k^{2}v_{0}^{2}}\Bigr].
\end{align}
\noindent At this point, we assume a large‑Péclet regime: for the most unstable wavenumber $k_{\max}$, $k_{\max}v_{0}\gg\sigma$.  Dropping the $k^{-2}v_{0}^{-2}$ terms contracts the condition above to
\begin{align}
  \rho_{0}\widehat F(k)>\frac{\sigma^{2}}{2}, \label{eq:dipole_firstCondition}\\[2pt]
  k^{2}v_{0}^{2}>\sigma^{2}\bigl[2\rho_{0}\widehat F(k)-\sigma^{2}\bigr]. \label{eq:dipole_secondCondition}
\end{align}

The condition in Equation~\eqref{eq:dipole_firstCondition} confirms that dipolar order can arise only after the focal cell's neighbourhood becomes radially inhomogeneous, whereas the condition in Equation~\eqref{eq:dipole_secondCondition} states that ballistic advection must outrun rotational diffusion, i.e.\ the angular Péclet number, $\mathrm{Pe}_{\theta}=kv_{0}/\sigma$, exceeds $\sqrt{2\rho_{0}\widehat F(k)-1}$.

\subsection{Second-order anisotropy: quadrupole instability}
\label{sec:side-patterns}
Front-back structure first appears when \(|m|+|n|=1\). To capture growth modes involving lateral patterns, one must allow for quadrupolar variations proportional to $\sin 2\Theta$, so that there are modulations at $\Theta=\pm\pi/2$. In terms of the Fourier indices $(m,n)$ of Equation~\eqref{eq:fixed-speed}, $\sin 2\Theta$ corresponds to the pair $(m,n)=(1,-1)$ and $(m,n)=(-1,1)$.
This defines the second anisotropic instability, $|m|+|n|=2$. Because the linear convective coupling links each mode to its four nearest neighbours, the smallest closed subset that contains, for instance, \((1,-1)\) is
\begin{equation}
  \mathscr B_2=\bigl\{(1,-1),\,(0,-1),\,(1,0)\bigr\}.
  \label{eq:B2}
\end{equation}
\subsubsection{Linearised system}
Writing
\(
   \mathbf g
   =(g_{1,-1},\,g_{0,-1},\,g_{1,0})^{\!\top}
\)
and defining, $\alpha_1=\lambda-\lambda_r(k)+2D_\theta$, $\alpha_d=\lambda-\lambda_r(k)+D_\theta$, and $\beta=k v_0/2$, Equation~\eqref{eq:fixed-speed} restricted to \(\mathscr B_2\) reads \(M_2(\lambda)\mathbf g=\mathbf 0\) with
\begin{equation}
  M_2(\lambda)=
  \begin{pmatrix}
    \alpha_1 & \beta & \beta \\
    -\beta   & \alpha_d & 0 \\
    -\beta   & 0 & \alpha_d
  \end{pmatrix}.
  \label{eq:M2}
\end{equation}
Non‑trivial solutions require
\(
  \det M_2(\lambda)=0
\). Taking the determinant and factorising gives
\begin{equation}
  (\lambda-\lambda_r+D_\theta)
    \bigl[(\lambda-\lambda_r)^2
          +3D_\theta(\lambda-\lambda_r)
          +2D_\theta^{2}+2\beta^{2}\bigr] = 0.
  \label{eq:detM2}
\end{equation}
The first factor reproduces the dipole growth rate from the previous section, {\(\lambda_{\mathrm{dip}}(k)\)}, and the quadratic bracket contains the new side‑pattern roots.

\subsubsection{Eigenvalues and instability threshold}
Solving the quadratic for the side-pattern roots from Equation~\eqref{eq:detM2} yields
\begin{equation}
    \lambda_{\mathrm{side},\pm}(k)
    =\lambda_r(k)-\tfrac32 D_\theta
     \;\pm\;\tfrac12\sqrt{\,D_\theta^{2}+8\beta^{2}\,}.
  \label{eq:lambda_side}
\end{equation}
These eigenvalues remain real, and oscillatory growth
does not arise in this block. Substituting \(\lambda_r(k)\) as before produces the threshold
\begin{equation}
  \rho_0\widehat F(k) > \frac{\sigma^2}{2} + \frac{1}{2k^2}\left(\frac{3\sigma^2}{2} - \sqrt{\frac{\sigma^4}{4} + 8\beta^2}\right)
  \label{eq:side_threshold_F}
\end{equation}
When this instability arises, the two modes, \((1,-1)\), and \((-1,1)\) combine into
\(
  \cos(\theta-\theta')\sin(\theta-\varphi)
  \propto\sin 2\Theta
\)
after setting \(\varphi=\theta+\Theta\).
Therefore a positive \(\lambda_{\text{side},+}\) produces lobes at
\(\Theta=\pm\pi/2\) in the neighbour distribution and, via the kernel
\(f_{\mathrm{ar}}(r,\Theta)\), drives a quadrupolar component
\(\propto\sin 2\Theta\) in the velocity-velocity correlation
\(C(r,\Theta,t)\). Finally, if \(\lambda_{\pm1}(k)\) is already positive, the dipole modes feed the quadrupole through the same \(\beta\)-coupling. That nonlinear reinforcement falls beyond the present linear analysis but is physically plausible.

\subsection{A minimal dipolar kernel that suppresses dipolar patterns}
\label{sec:minimal_anisotropy}
A converse question to the one that have been asked so far, is whether observing a kernel that looks largely isotropic must mean that the interaction kernel must be isotropic. The answer to this question is no, and, as will be shown in this section, the reason lies in a strict hierarchy of angular instabilities. Before embarking on this analysis, it is worth noting that clean side-by-side patterns, as instabilities of the linearised system, cannot occur with a radially isotropic interaction force alone. This is because the quadrupolar branch responsible for the lateral lobes in Equation~\eqref{eq:lambda_side} becomes positive only after the density drive has surpassed its $3D_{\theta}/2$ threshold, whereas the dipolar branch requires only a threshold of $D_{\theta}$. Now, to explore how isotropically-looking patterns can emerge from a non-isotropic interaction force, we consider a dipolar (\textit{i.e.} front-back) interaction force, $f_{\mathrm{ar}}$. The arguments in this section will show conditions under which such a mechanism can suppress dipolar instabilities that would be expected normally with an isotropic kernel. 

\subsubsection{Kernel with a single dipole harmonic and its Fourier transform}
\label{sec:dipole-kernel}
Consider an interaction kernel that contains only the monopole and a single dipole harmonic,
\begin{equation}
  f_{\!\text{ar}}(r,\phi)=F_0(r)+\varepsilon\,F_1(r)\cos\phi,
  \qquad 0<|\varepsilon|\ll1.
  \label{eq:kernel-dipole}
\end{equation}
The small parameter~$\varepsilon$ measures the strength of a front-back bias relative to the isotropic attraction. Consider again the expansion of perturbations of $\tilde P$ in a Fourier-Bessel basis (as done in Sections~\ref{sec:first-order-modes} and~\ref{sec:side-patterns}. As before, the purely radial contribution appears in every block and gives a
baseline growth rate
\(\lambda_r(k)=-k^{2}D_t+\rho_0k^{2}\widehat F(k)\).
Projecting this common term onto a monopole,
\(p=0\), \textit{i.e.}, taking the inner product with
\(\mathrm e^{\mathrm i0\phi}\), yields the familiar leading eigenvalue. This term, being proportional to the zeroth harmonic \(F_0\), is then inherited by every angular sector of the linear
operator.

Introducing the dipolar kernel in Equation~\eqref{eq:kernel-dipole} now adds a first angular harmonic but no higher ones.
Writing \(\cos\phi=\frac12(\mathrm e^{\mathrm i\phi}+\mathrm
e^{-\mathrm i\phi})\) shows that the only new Fourier coefficients are
\(F_{\pm1}=\varepsilon F_1/2\).
When the interaction matrix is projected onto
\(p=\pm1\), \textit{i.e.}, onto the dipoles, this additional coefficient
contributes a diagonal shift
\[
  \alpha(k)=\frac{\rho_0k^{2}\varepsilon}{2}\,\widehat F_1(k),
\]
where, analogously to the sections before, $\widehat F_1(k)\equiv2\pi\!\int_0^\infty rF_1(r)J_1(kr)\,dr$. If \(\varepsilon\widehat F_1 < 0\), then
\(\alpha(k)<0\). In this case the dipole growth
rate is depressed. \textit{Vice versa}, a positive value reinforces it.

\subsubsection{Updated dipole and quadrupole eigenvalue blocks}
Since for each radial wave‑number~$k$, the isotropic sector is unchanged, but the dipole block carries
an extra diagonal entry $+\alpha(k)$, it follows that the  eigenvalue corresponding to the dipole block is now defined by
\begin{equation}
  \;
  \lambda_{\text{dip}}(k)=\lambda_r^{(0)}(k)+\alpha(k)-D_\theta
  \;+\;
  \frac{\beta^{2}}{\lambda_r^{(0)}(k)+D_\theta}.\;
  \label{eq:lambda-dipole}
\end{equation}
Recall that the quadrupole (side) matrix, $M_2(\lambda)$, of
Section~\ref{sec:side-patterns} does not involve the $p=\pm1$ modes, so its eigenvalues are unaffected and are given by Equation~\eqref{eq:lambda_side}.

\subsubsection{Instability conditions for a suppressed dipolar pattern}
To find instabilities such that an emergent dipole pattern is suppressed, one must seek wavenumbers for which
\(\lambda_{\text{side},+}(k)>0\) and
\(\lambda_{\text{dip}}(k)<0\).
Using Equations~\eqref{eq:lambda_side}~and~\eqref{eq:lambda-dipole}, it is sufficient that
\begin{equation}
  D_\theta < \lambda_r^{(0)}(k) < D_\theta-\alpha(k).
  \label{eq:window}
\end{equation}
Equation~\eqref{eq:window} can hold only if
\(\alpha(k)<0\), i.e.\ $\varepsilon\widehat F_1(k)<0$.
Hence a moderately repulsive dipole component ($\varepsilon F_1<0$) depresses the dipole growth rate below zero while the isotropic drive remains large enough to trigger the quadrupole. Translated to microscopic parameters, the coexistence band for a suppressed dipole pattern instability to arise is defined by
\begin{equation}
  \varepsilon \widehat F_1(k) < -2\widehat F_0(k) + \frac{\sigma^2}{\rho_0}\left(1 + \frac{1}{k^2}\right). 
  \label{eq:micro-window}
\end{equation}
Increasing $|\varepsilon|$, \textit{i.e.} strengthening the dipolar repulsion, widens the parameter window in which dipole pattern instabilities can be suppressed. Put together, anisotropy in the interaction kernel is required, but its sign and magnitude decide whether the observable correlation field ends up anisotropic, and also which multipole wins.  A dipolar kernel
can, counter‑intuitively, produce a quadrupolar correlation
pattern that is not hard‑wired in the force itself.

\subsection{Numerical validation}
To test the instability predictions of Sections~\ref{sec:first-order-modes}-\ref{sec:minimal_anisotropy} on empirical trajectories, a classic framework with an attraction-repulsion kernel~\cite{bertozzi2006self} is used as a specific example of an SPP model in the form of Equation~\eqref{eq:dynamics}. We compare the resulting stationary correlation PDE, and direct simulations with the resulting correlation fields. The model is given by
\begin{equation}
  S(\mathbf v_i)=\alpha-\beta\lvert\mathbf v_i\rvert, \qquad
  f_{\mathrm{ar}}(r)= \mathcal M(r), \qquad
  \mathcal M(r)= \frac{C_R}{L_R}e^{-r/L_R}-\frac{C_A}{L_A}e^{-r/L_A},
  \label{eq:dorsogna_short}
\end{equation}
with $\alpha=\beta$ so that the preferred speed is unity. Crucially, the attraction-repulsion force is strictly radial. We simulate $N$ agents in a periodic square of side $L=70$ using an Euler-Maruyama scheme with $\Delta t=0.02$ for $2000$ time steps, using the first $500$ time steps as burn-in. The number of agents is chosen to achieve the target density $\rho_0=N/L^2$ in each regime. Empirical correlations are measured in each focal particle’s frame:
\[
C_{\mathrm{emp}}(r,\theta)=
\big\langle\,\hat{\mathbf v}_i\!\cdot\!\hat{\mathbf v}_j\,\big\rangle_{\,j:\,
r_{ij}\in[r,r+\Delta r],\ \theta_{ij}\in[\theta,\theta+\Delta\theta]},
\]
where $\theta_{ij}$ is the polar angle of $\mathbf r_{ij}$ relative to $\hat{\mathbf v}_i$ and neighbours are binned on a polar grid $(r,\theta)$. For the isotropic part of the force we use a Morse form \(F_0(r)=\mathcal{M}(r)\), and for the anisotropic dipole we choose  \(F_1(r,\phi)\) as before, so that
\[
\widehat F_0(k)=2\pi\!\left[
\frac{C_R/L_R^2}{(L_R^{-2}+k^2)^{3/2}}
-\frac{C_A/L_A^2}{(L_A^{-2}+k^2)^{3/2}}\right],\qquad
\widehat F_1(k)=2\pi A_1\,\frac{k}{(L_1^{-2}+k^2)^{3/2}}.
\]
This is used to compute the dispersion curves for the $\ell=0,1,2$ branches and identify the most-unstable wavenumber $k_\ast$ used to select the branch for each regime. For the PDE comparison we solve, in angular Fourier components, the steady radial boundary-value problems implied by Equation~\eqref{eq:C_evolution}. Each mode $C_m(r)$ satisfies a second-order ODE on $r\in[0,R]$ with $R=L/2$, driven by a spatially constant source, $S_m(r)$. Boundary conditions are regularity at $r=0$ (Neumann for $m=0$, $C_m(0)=0$ for $m\ge1$) and $C_m(R)=0$. We discretise with a uniform radial grid using second-order centered differences and solve the resulting banded linear systems directly. For the quadrupolar column we superpose the $m=0$ and $m=2$ solutions with the mixing ratio $u_2/u_0$ evaluated at $k_\ast$ from the side branch eigenvector; other columns display a single mode ($m=0$ or $m=1$). Fields are normalized to unit maximum absolute value before plotting in the co-moving frame.
\begin{figure}[h!]
    \centering
    \includegraphics[width=\linewidth]{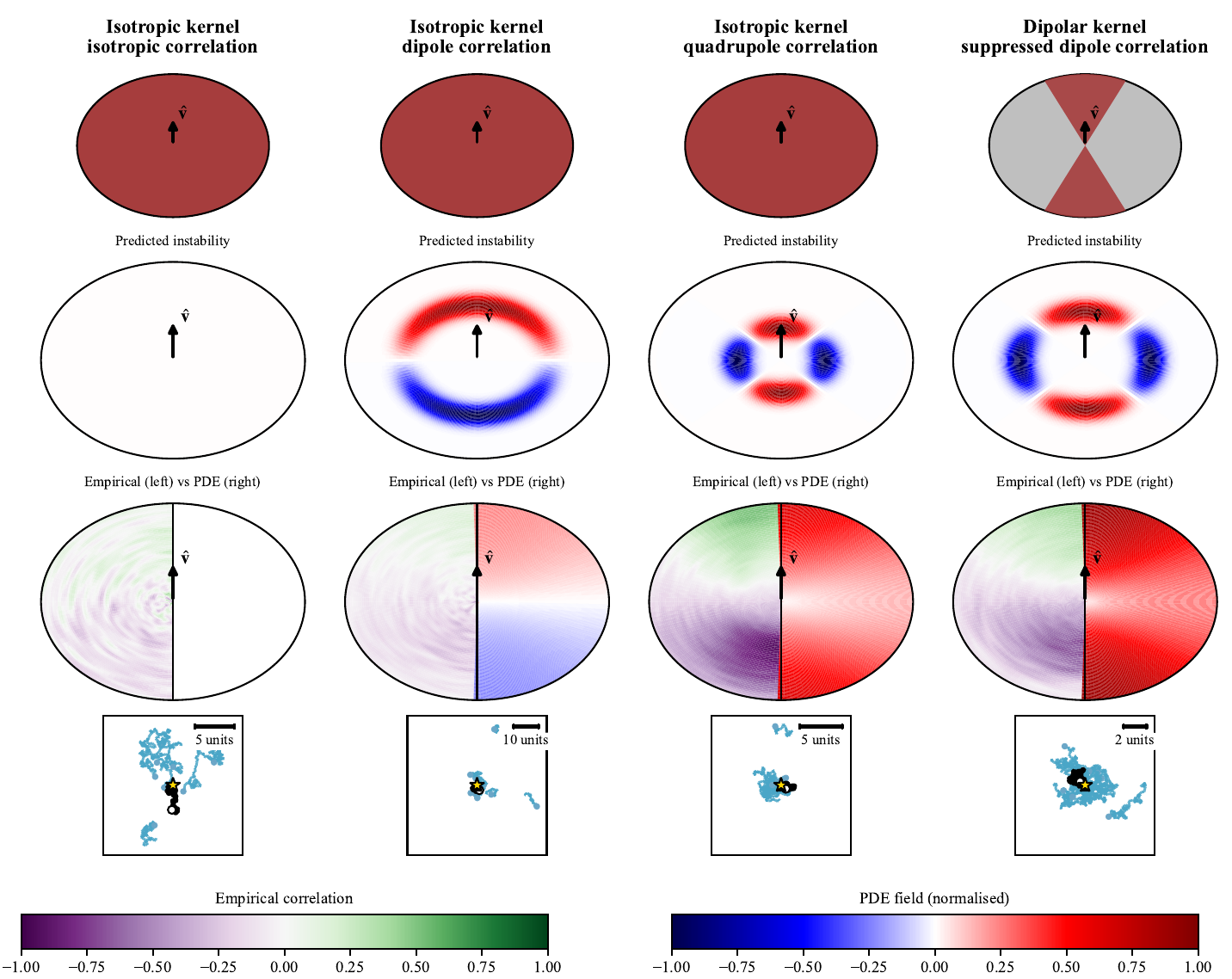}
    \caption{Numerical validation of the instability theory. Top row: schematics of the interaction kernels. Second row: for each regime the most unstable harmonic predicted by the dispersion analysis as $C(r,\theta)$ on a ring. Third row: left semicircles show the empirical correlation field from simulations and right semicircles show the stationary solution of the linear PDE evaluated with the same parameters; the arrow indicates the focal heading $\hat{\mathbf v}$. Bottom: representative focal and neighbor trajectories with scale bars. Simulation parameters are listed in Table~\ref{tab:parameters}. Note that the PDE predictions align well with the empirical dipole pattern in the second column, but fail to capture the exact structure in the quadrupole and suppression regimes, highlighting that leading-order linear instabilities are not always a reliable proxy for the fully developed PDE solutions.}
    \label{fig:trajectories}
\end{figure}

We set $L_R=0.5$, $L_A=2.0$, $C_R=1.0$, $C_A=0.95$, and $v_0=1$. Further, the parameter choices listed in Table~\ref{tab:parameters} are designed to isolate the behaviours discussed in Sections~\ref{sec:first-order-modes}-\ref{sec:minimal_anisotropy}. Figure~\ref{fig:trajectories} shows the computed correlations and empirical trajectories for each of these parameter regimes. In the isotropic kernel at low density and high noise, the correlation field is nearly flat, consistent with no unstable modes. As density increases, the dipole branch crosses zero, producing a faint but detectable front-back asymmetry. At still higher density the quadrupolar branch dominates, leading to side lobes visible in the correlation field. Finally, when a weak anisotropic dipolar term is added to the kernel, the dipole instability is suppressed: the empirical correlation is reduced in amplitude, consistent with the predicted suppression window. 

In Figure~\ref{fig:trajectories}, the agreement between the predicted linear mode of the PDE and empirical correlation is clearest in the isotropic-dipole case, where the PDE prediction reproduces the front-back bias seen empirically. In contrast, for the isotropic-quadrupole and dipole-suppression regimes the linearised PDE fields do not resemble the full empirical correlations. This mismatch is expected as the PDE fields are constructed from the leading linear instability alone, whereas the empirical correlations reflect nonlinear saturation, harmonic mixing, and finite-density effects. Linear theory is therefore a good proxy only when a single branch is near threshold, but not when multiple instabilities interact or are strongly subcritical. Nevertheless, the qualitative features of growing magnitude of the growth coefficient and the subsequent suppression are recapitulated by the linear theory in Figure~\ref{fig:trajectories}.
\begin{table}
\begin{tabular}{lcccc}
\toprule
Case & Kernel & \(\rho_0\) & \(\sigma\) & Extra parameters \\
\midrule
Isotropic / isotropic correlation & \(F_0\) & 0.08 & 0.90 & --- \\
Isotropic / dipolar correlation   & \(F_0\) & 0.35 & 1.0 & --- \\
Isotropic / quadrupolar correlation & \(F_0\) & 0.90 & 0.45 & --- \\
Dipolar kernel / suppressed dipole & \(F_0+F_1\) & 0.55 & 0.80 &
\(\varepsilon=-0.50,\;A_1=1.0,\;L_1=1.0\) \\
\bottomrule
\end{tabular}
\caption{Parameter values used for the simulation of interacting agents according to the model in Equation~\eqref{eq:dorsogna_short}.}
\label{tab:parameters}
\end{table}

\begin{figure}[h!]
    \centering
    \includegraphics[width=0.5\linewidth]{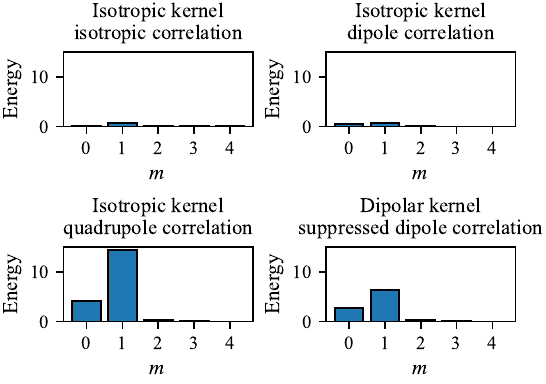}
    \caption{Angular-harmonic decomposition of empirical correlations. For each regime, the empirical correlation field from Figure~\ref{fig:trajectories} is decomposed orthogonally into angular harmonics.}
    \label{fig:ell_modes}
\end{figure}

To verify the extent to which the linearly predicted unstable modes are present in the empirical correlation fields that are computed for each of the regimes, the empirical correlation fields from Figure~\ref{fig:trajectories} are decomposed orthogonally into angular harmonics, such that
\begin{equation*}
    C(r,\theta)=a_0(r)+\sum_{m\ge1}a_m(r)\cos(m\theta).
\end{equation*}
The bar plots in Figure~\ref{fig:ell_modes} show the contributions, $E_m$, of each harmonic, which are defined as
\begin{equation}
    E_0=2\pi\int a_0(r)^2\,r\,dr,\qquad E_m=\pi\int a_m(r)^2\,r\,dr\ (m\ge1).
\end{equation}
Identifying the contributions of each harmonic isolates which harmonic is present in each regime. The bar plots of Figure~\ref{fig:ell_modes} confirm the pattern shown by the sequence of instabilities discussed in Figure~\ref{fig:trajectories}. In the regime predicted to host a dipole instability, the $m=1$ contribution remains subdominant, consistent with the linear growth rate being weak at the chosen parameters. In the quadrupole regime, higher modes contribute alongside $m=2$, showing that the empirical correlation saturates as a mixture rather than a pure eigenmode. The dipole-suppression case shows that the dominant eigenmodes are suppressed relative to the quadrupolar correlation regime.

\section{Effect of velocity alignment on emergent correlations}
\label{sec:alignment}
Sections~\ref{sec:forward} and \ref{sec:emergence} showed that isotropic attraction-repulsion kernels alone can generate emergent anisotropic correlations. While this isolates the role of pure attraction and repulsion anisotropy, many examples of interacting agents in biology include velocity alignment explicitly in response to neighbours~\cite{te_Vrugt_2025}. To explore how such an extra model term influences the emergence of correlations on top of attraction-repulsion dynamics, here we incorporate velocity alignment in the simplest, linear, form compatible with the model used above. We extend the single-particle dynamics in Equation~\eqref{eq:kinematics} by adding an additional force, $F_{\mathrm{align}}$  that rotates the heading of cell $i$ toward the heading of $j$,
\begin{equation}
  F_{\mathrm{align}} = \frac{v_0}{N}\sum_{j\neq i}
  f_{al}(r_{ij},\varphi_{ij})\,(I-\hat u_i\hat u_i^\top)\hat u_j,
  \label{eq:al-law}
\end{equation}
where $\hat u_i=v_i/|v_i|$. The projector, $(I-\hat u_i\hat u_i^\top)$, removes the component along $\hat u_i$, so alignment only rotates headings and preserves speed. For simplicity, we parameterise the distance- and angle-dependent alignment density, $f_{\text{al}}$, by assuming that it can be written as
\begin{equation}
  f_{al}(r,\theta)=\eta_0 F_0(r)+\eta_1\,\varepsilon F_1(r)\cos\theta,
  \label{eq:al-kernel}
\end{equation}
\textit{i.e.} so that the Bessel transforms computed for the alignment kernel will be proportional to those of the attraction-repulsion kernel, $f_{\text{ar}}$. The alignment drift in Equation~\eqref{eq:al-law} will now contribute to the Fokker-Planck equation, 
\begin{equation}
  \partial_t C \;=\; -v_0\,\partial_r C
  + D_r\!\left(\partial_r^2 + r^{-1}\partial_r\right)C
  + \frac{D_\theta}{r^2}\partial_\theta^2 C
  - 2\gamma\,C \;+\; f_{ar} + f_{al},
  \label{eq:C-with-alignment}
\end{equation}
see Appendix~\ref{app:alignmentContributions} for details. Thus, alignment enters additively, with the same harmonic content as $f_{ar}$, and the Fourier-Bessel block structure of Section~\ref{sec:emergence} is unchanged. Alignment changes only by having the first harmonic in $f_{al}$ contributing a diagonal shift to the dipole block, exactly as a dipolar kernel, $f_{\text{ar}}$ does in Equation~\eqref{eq:kernel-dipole}, while leaving the side matrix 
of Section~\ref{sec:side-patterns} unaltered at leading order. Defining $\widehat{A}_1(k)$ as
\[
  \widehat{A}_1(k)=2\pi\int_0^\infty r\,\big( F_1(r)\big)J_0(kr)\,dr,
\]
the dipole eigenvalue becomes
\begin{equation}
  \lambda_{\mathrm{dip}}(k)
  \;=\; \lambda_r^{(0)}(k) \;+\; \alpha_{ar,1}(k)\;+\;
  \alpha_{al,1}(k) \;-\; D_\theta \;+\;
  \frac{\beta^2}{\lambda_r^{(0)}(k)+D_\theta},
  \label{eq:dip-with-al}
\end{equation}
where
\begin{equation}
  \alpha_{al,1}(k)=\frac{\rho_0 k^2}{2}\eta_1 \varepsilon\,\widehat{A}_1(k),
  \qquad
  \alpha_{ar,1}(k)=\frac{\rho_0 k^2}{2}\,\varepsilon\,\widehat{F}_1(k),
  \label{eq:alphas}
\end{equation}
and $\beta=kv_0/2$ as before. The side branch eigenvalues remain the same, as no new terms enter. Two consequences follow immediately. First, the thresholds needed to induce a dipole instability shift, as any $\eta_1\neq 0$ simply replaces the dipole shift that occurs due to attraction-repulsion by $\alpha_{ar,1}\!+\alpha_{al,1}$ in Equation~\eqref{eq:dip-with-al}. A positive alignment term given by $\eta_1$ therefore raises $\lambda_{\mathrm{dip}}$ across $0$ at lower $v_0$ or density, whereas a negative $\eta_1$ (\textit{i.e.}, anti-alignment)  depresses it. Second, the parameter window in which dipolar instabilities are depressed can be either created or erased. Because the side matrix, $M_2$, is unchanged, the inequality that guarantees this window is still
\begin{equation}
  D_\theta \;<\; \lambda_r^{(0)}(k) \;<\; D_\theta - \big(\alpha_{ar,1}(k)+\alpha_{al,1}(k)\big),
  \label{eq:clean-side-with-al}
\end{equation}
\textit{cf.} Equation~\eqref{eq:window} with $\alpha\mapsto\alpha_{ar,1}+\alpha_{al,1}$. Thus, anti-alignment in the front-back sector ($\eta_1<0$) widens this window even when $f_{ar}$ is isotropic, while pro-alignment narrows or destroys it. In this minimal extension, alignment can selectively modulate the dipole instability with no change to the side matrix, and the final resulting pattern hinges on the sign and magnitude of $\alpha_{al,1}(k)$.

\begin{figure}
    \centering
    \includegraphics[width=0.66\linewidth]{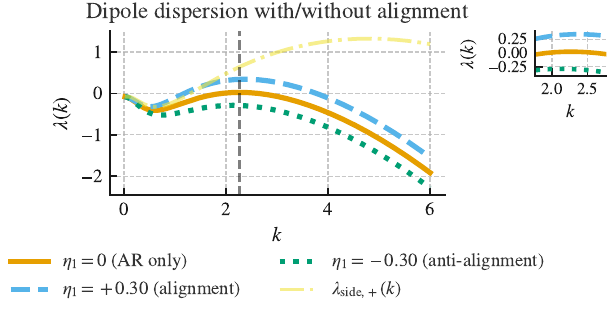}
    \caption{Dispersion curves, $\lambda(k)$, for the dipole branch with no alignment $(\eta_1=0)$, with alignment $(\eta_1>0)$ and anti-alignment $(\eta_1<0)$, together with the side branch $\lambda_{\text{side},+}(k)$. Alignment shifts only the dipole curve by $\alpha_{\mathrm{al},1}(k)$ in Equation~\eqref{eq:dip-with-al}, while $\lambda_{\text{side},+}(k)$ is unaffected at leading order. Inset: zoom near $k^\ast$.  }
    \label{fig:dipole-dispersion}
\end{figure}

As an illustration, Figure~\ref{fig:dipole-dispersion} shows the dispersion relation for the same parameters as the suppressed dipole regime in Figure~\ref{fig:trajectories} with a weak dipolar component in the attraction-repulsion kernel ($\epsilon = -0.5$, $A_1 = 1.0$, $L_1 = 1.0$). At these parameters, the Morse attraction–repulsion kernel alone produced an unstable quadrupole instability as well as an unstable dipole instability (with a smaller growth rate). We now introduce a separate dipolar alignment term, with amplitude~$\eta_1$. As discussed previously, this term can selectively shift the dipole eigenvalue without affecting the quadrupole growth rate. We therefore explore $\eta_1 \in \{0,\, +0.3,\, -0.3\}$, corresponding respectively to: no alignment, alignment, and anti-alignment, while all other parameters remain fixed. For $\eta_1=0$, the dispersion relation $\lambda_{\mathrm{dip}}(k)$ exhibits a maximum of zero near $k\simeq2.1$, indicating a critical point for an emergent dipole instability, as before. For $\eta_1>0$ (alignment), $\alpha_{\mathrm{al},1}(k)>0$ and the dipole curve shifts upward, slightly enhancing this mode. For $\eta_1<0$ (anti-alignment), $\alpha_{\mathrm{al},1}(k)<0$ and the dipole curve shifts downward while the side branch $\lambda_{\mathrm{side},+}(k)$ remains unchanged, thereby creating a distinct band of $k$ for which
\[
\lambda_{\mathrm{dip}}(k) < 0 < \lambda_{\mathrm{side},+}(k),
\]
corresponding to a parameter regime in which the dipole instability is even further suppressed. Put together, this demonstrates that an alignment term can selectively dampen or strengthen the dipole instability while leaving higher modes largely unchanged, thus fundamentally changing the linear stability properties of the emergent correlation field.

\section{Discussion}
By deriving a kinetic theory-based closed-form PDE from a canonical SPP model, this work establishes conditions under which intrinsic directional biases can be reliably distinguished from emerging anisotropy in the velocity correlations between two interacting particles. A linear stability analysis revealed that front-back or lateral biases may arise spontaneously and various kinetic mechanisms can contribute to the emergence or suppression of such instabilities. This work therefore complements recent developments in modelling of active matter~\cite{JiangLi2017Iini, brunton2016discovering, messenger2022learning} providing an explicit analytical framework for the angular structure of correlations. These findings underscore the importance of ruling out alternative explanations before concluding that directional forces drive observed anisotropic behaviors. In fact, the theory developed in this paper can be used as the basis for statistical testing whether observed velocity correlations in collectively migrating agents are due to an inherent anisotropy in the interaction kernel. This is because, the spectrum of the operator that dictates the evolution of the correlation field specifies which harmonics are expected to be stable under isotropic interactions. Angular decomposition of observed correlations would identify which modes are present. Using a noise model, the magnitude of such modes can be tested against the null model of what can be expected given an isotropic kernel. Developing such tests is an obvious next step for applications to experimental data.

This work can be extended in several ways. First, a formal analysis of nonlinear feedback mechanisms among kinetic processes may elucidate more complex collective behaviors beyond linear stability predictions. Second, integrating this kinetic-theory approach into a fully Bayesian inference framework would enable rigorous quantification of uncertainty and facilitate inference of the true underlying interaction kernels. Finally, employing additional biological data, such as tracking cell-cell contacts, cell morphology, or organelle dynamics through fluorescent microscopy, could provide deeper mechanistic insights and suggest plausible biological mechanisms giving rise to the observed interaction laws. 

As a final comment, the multipole‐instability mechanism discussed in this paper is directly analogous to the classical Turing instability in reaction-diffusion systems.  There, an isotropic, slowly diffusing `activator' species and a more rapidly diffusing `inhibitor' species select a finite band of spatial wavenumbers. This interaction produces stationary stripes or spots, despite both species and their diffusion being isotropic \cite{turing1952chemical}.  In the Fokker-Planck equation framework, different angular harmonics of an isotropic interaction kernel play the role of distinct `species'.  The monopole (isotropic drive) supplies the baseline growth of correlations, while the dipole harmonic functions as an angular `inhibitor', selectively damping front-back modes in a wavenumber‐dependent fashion.  When this dipolar sector is sufficiently suppressed, the quadrupole mode becomes unstable over a finite band of \(k\), yielding clear side‐aligned correlation `lanes'.  Hence, anisotropic patterns in the two‐point velocity correlation arise not from explicit breaking of rotational symmetry in the forces, but from the interplay of isotropic interaction modes and angular mixing.

\clearpage
\subsubsection*{Competing interests}
I declare I have no competing interests.

\subsubsection*{Acknowledgements}
I would like to thank the Foulkes Foundation for funding. I would also like to thank Professor Ruth Baker for our discussions inspiring this work. 


\newpage
\bibliography{references, manual_refs}
\bibliographystyle{unsrt}

\newpage
\appendix
\setcounter{equation}{0}\renewcommand\theequation{A\arabic{equation}}

\appendix

\section{From the \texorpdfstring{$N$}{N}-body Fokker-Planck equation to a pair equation}
\label{app:pair_fp_full}
We begin with the microscopic equations of motion for $N$ interacting self-propelled particles:
\begin{align}
\dot{\mathbf{x}}_i &= \mathbf{v}_i, \\
\dot{\mathbf{v}}_i &= S(\mathbf{v}_i)\mathbf{v}_i + \frac{1}{N} \sum_{j \neq i} f_{\text{ar}}(r_{ij}, \phi_{ij}) \hat{\mathbf{r}}_{ij} + \sigma \boldsymbol{\xi}_i(t),
\end{align}
where $\hat{\mathbf{r}}_{ij} = \frac{\mathbf{x}_j - \mathbf{x}_i}{\norm{\mathbf{x}_j - \mathbf{x}_i}}$ and $\phi_{ij}$ is the angle between $\mathbf{v}_i$ and $\mathbf{x}_j - \mathbf{x}_i$. This stochastic system corresponds to the $N$-particle distribution function $f_N(\mathbf{x}_1, \mathbf{v}_1, \dots, \mathbf{x}_N, \mathbf{v}_N, t)$. According to the Itô-Kolmogorov forward equation, this density evolves under the $N$-body Fokker-Planck equation:
\begin{align}
\frac{\partial f_N}{\partial t} &+ \sum_{i=1}^N \nabla_{\mathbf{x}_i} \cdot (\mathbf{v}_i f_N)
+ \sum_{i=1}^N \nabla_{\mathbf{v}_i} \cdot \left[
\left( S(\mathbf{v}_i)\mathbf{v}_i + \frac{1}{N} \sum_{j \neq i} f_{\text{ar}}(r_{ij}, \phi_{ij}) \hat{\mathbf{r}}_{ij} \right) f_N
\right] \nonumber \\
&= \frac{\sigma^2}{2} \sum_{i=1}^N \Delta_{\mathbf{v}_i} f_N.
\end{align}

\subsection*{Reduced Distributions}
We define the one-particle and two-particle reduced distribution functions as marginals of $f_N$:
\begin{align}
f_1(\mathbf{x}_1, \mathbf{v}_1, t) &= \int f_N \, d\mathbf{x}_2 d\mathbf{v}_2 \dots d\mathbf{x}_N d\mathbf{v}_N, \\
f_2(\mathbf{x}_1, \mathbf{v}_1, \mathbf{x}_2, \mathbf{v}_2, t) &= \int f_N \, d\mathbf{x}_3 d\mathbf{v}_3 \dots d\mathbf{x}_N d\mathbf{v}_N.
\end{align}
Our goal is to derive an evolution equation for $f_2$ by integrating the $N$-body equation over particles $3$ to $N$.

\subsection*{BBGKY Equation for $f_2$}
Integrating the $N$-body equation over particles $3, \dots, N$ yields:
\begin{align}
\frac{\partial f_2}{\partial t} &+ \nabla_{\mathbf{x}_1} \cdot (\mathbf{v}_1 f_2) + \nabla_{\mathbf{x}_2} \cdot (\mathbf{v}_2 f_2) \nonumber \\
&+ \nabla_{\mathbf{v}_1} \cdot \left[
S(\mathbf{v}_1)\mathbf{v}_1 f_2 + \frac{1}{N} f_{\text{ar}}(r_{12}, \phi_{12}) \hat{\mathbf{r}}_{12} f_2 + \frac{1}{N} \sum_{k \geq 3} \int f_{\text{ar}}(r_{1k}, \phi_{1k}) \hat{\mathbf{r}}_{1k} f_3 \, d\mathbf{x}_3 d\mathbf{v}_3
\right] \nonumber \\
&+ \nabla_{\mathbf{v}_2} \cdot \left[
S(\mathbf{v}_2)\mathbf{v}_2 f_2 + \frac{1}{N} f_{\text{ar}}(r_{21}, \phi_{21}) \hat{\mathbf{r}}_{21} f_2 + \frac{1}{N} \sum_{k \geq 3} \int f_{\text{ar}}(r_{2k}, \phi_{2k}) \hat{\mathbf{r}}_{2k} f_3 \, d\mathbf{x}_3 d\mathbf{v}_3
\right] \nonumber \\
&= \frac{\sigma^2}{2} \left( \Delta_{\mathbf{v}_1} + \Delta_{\mathbf{v}_2} \right) f_2.
\end{align}

\subsection*{Mean-field (Molecular Chaos) Closure}
We now close the hierarchy by approximating the three-particle distribution function using a product of lower-order distributions:
\begin{align}
f_3(\mathbf{x}_1, \mathbf{v}_1, \mathbf{x}_2, \mathbf{v}_2, \mathbf{x}_3, \mathbf{v}_3) \approx f_2(\mathbf{x}_1, \mathbf{v}_1, \mathbf{x}_2, \mathbf{v}_2) f_1(\mathbf{x}_3, \mathbf{v}_3).
\end{align}
Using this, we obtain an approximate expression for the additional interaction term:
\begin{align}
\frac{1}{N} \sum_{k \geq 3} \int f_{\text{ar}}(r_{1k}, \phi_{1k}) \hat{\mathbf{r}}_{1k} f_3 \, d\mathbf{x}_3 d\mathbf{v}_3
\approx \frac{N-2}{N} f_2 \int f_{\text{ar}}(r_{13}, \phi_{13}) \hat{\mathbf{r}}_{13} f_1(\mathbf{x}_3, \mathbf{v}_3) \, d\mathbf{x}_3 d\mathbf{v}_3.
\end{align}
For large $N$, $\frac{N-2}{N} \approx 1$, so we simplify this further.

\subsection*{Pair-Density Representation}
Define the pair-probability density in terms of relative coordinates:
\begin{align}
P(\mathbf{r}, \mathbf{v}, \mathbf{v}', t) = \left\langle \sum_{i \ne j} \delta(\mathbf{r} - \mathbf{x}_j + \mathbf{x}_i) \delta(\mathbf{v} - \mathbf{v}_i) \delta(\mathbf{v}' - \mathbf{v}_j) \right\rangle.
\end{align}
This is a symmetrized version of $f_2$ written in relative coordinates $\mathbf{r} = \mathbf{x}_j - \mathbf{x}_i$ and velocities $\mathbf{v}, \mathbf{v}'$.

\subsection*{Fokker-Planck Equation}
We now obtain the evolution equation for $P$ by transforming the equation for $f_2$:
\begin{align}
\label{eq:Fokker-Planck_appendix}
\partial_t P + (\mathbf{v}' - \mathbf{v}) \cdot \nabla_{\mathbf{r}} P + \mathcal{L}_{\mathbf{v}}[P] + \mathcal{L}_{\mathbf{v}'}[P] = \frac{\sigma^2}{2} \left( \Delta_{\mathbf{v}} + \Delta_{\mathbf{v}'} \right) P,
\end{align}
where the operator $\mathcal{L}_{\mathbf{v}}[P]$ accounts for self-propulsion and interaction drift:
\begin{align}
\mathcal{L}_{\mathbf{v}}[P] = \nabla_{\mathbf{v}} \cdot \left( -S(\mathbf{v})\mathbf{v} - \int f_{\text{ar}}(r, \phi(\mathbf{v}, \mathbf{r})) \hat{\mathbf{r}} \, P(\mathbf{r}, \mathbf{v}, \mathbf{v}', t) \, d\mathbf{r} d\mathbf{v}' \right).
\end{align}
This completes the BBGKY reduction to a Fokker-Planck equation for the pair-probability density.

\newpage
\section{Conditional correlation field \texorpdfstring{$C(r,\theta,t)$}{C(r,theta,t)}}
\label{app:cond_corr_full}

The objective of this section is to extract a closed evolution equation for the
correlation field defined in polar variables centred on a focal
particle. 

\subsection*{Geometry of the co-moving frame}
For a chosen focal particle $i$ align the local $x$-axis with
$\mathbf v_i/v_0=\hat{\mathbf u}=(1,0)$.
For its neighbour $j$ set
\begin{equation}
  r:=r_{ij}=\|\mathbf x_j-\mathbf x_i\|,\qquad
  \theta:=\angle(\hat{\mathbf u},\mathbf x_j-\mathbf x_i).
\end{equation}
Write the headings as
$\hat{\mathbf u}=(1,0)$ and
$\hat{\mathbf u}'=(\cos\psi,\sin\psi)$ with
$\psi=\varphi-\varphi'$.  In this frame $\mathbf v=v_0\hat{\mathbf u}$ and
$\mathbf v'=v_0\hat{\mathbf u}'$.

\subsection*{Pair density, distribution function and correlation field}
Define the following quantities,
\begin{align}
  n(r,\theta,t) &:= \iint P\,d\varphi\,d\varphi',\\
  g(r,\theta,t) &:= n/\rho,\\
  C(r,\theta,t) &:= \frac{1}{n}\iint(\hat{\mathbf u}\!\cdot\!\hat{\mathbf u}')P\,d\varphi\,d\varphi'.\label{eq:defC_full}
\end{align}
Here \(\rho\) is the bulk (spatially averaged) particle number density, \textit{i.e.},
\[
  \rho \;=\;
  \int f_1(\mathbf x,\mathbf v,t)\,d\mathbf v.
\]
Dividing the pair density \(n(r,\theta,t)\) by this constant scale makes the pair‑distribution function \(g=n/\rho\) dimensionless. Because $\hat{\mathbf u}\!\cdot\!\hat{\mathbf u}'=\cos\psi\in[-1,1]$ it immediately follows that $-1\le C\le1$.

\subsection*{Step 1: angular integral identity}
Using $\partial_\varphi^2\cos\psi=-\cos\psi$ one finds
\begin{equation}
  \int_0^{2\pi}\!\int_0^{2\pi}\cos\psi\,P\,d\varphi\,d\varphi'=nC.
  \label{eq:angular_identity}
\end{equation}
This identity will be invoked in later steps.

\subsection*{Step 2: multiplying the Fokker-Planck equation by \texorpdfstring{$\cos\psi$}{cos(psi)}}
Insert $\cos\psi$ into Equation~\eqref{eq:Fokker-Planck_appendix}, integrate over
$\varphi,\varphi'$ and treat each term separately.
We catalogue them as (i)-(v) for future reference.

\paragraph*{(i) Time derivative.}
Trivial: $\partial_t(nC)$.

\paragraph*{(ii) Self-propulsion term.}
For the deterministic (drift) part of the motion, it can be noted that, because $\mathbf v'-\mathbf v=v_0(\hat{\mathbf u}'-\hat{\mathbf u})$ and
$\hat{\mathbf u}=(1,0)$,
\begin{align}
  &\iint \cos\psi\,(\mathbf v'-\mathbf v)\!\cdot\!\nabla_{\!r}P\,d\varphi\,d\varphi'
  =v_0\,\partial_r(nC).\label{eq:free_flight_raw}
\end{align}
However, since it restricts to drift alone, the calculation above keeps only the first so-called Kramers-Moyal coefficient of the term related to self-propulation, so it captures the ballistic drift of the mean separation but omits the higher‑order moments that generate effective diffusion after the headings decorrelate. Kramers-Moyal theory \cite{Kramers1940, Moyal1949, cercignani_book} expresses the infinitesimal propagator of any Markov process as an infinite series of differential operators.  The first coefficient gives drift, and the second gives diffusion. By retaining the second coefficient in Step~4 we recover the radial and angular Laplacians that encode the random‑walk broadening of the particle pair—physics that would be missed if one stopped at the drift level alone. In Step~4 we therefore revisit exactly the same term, rewrite it as a divergence of a flux, and carry out a systematic Kramers-Moyal expansion.

\paragraph*{(iii) Drift operators $\mathcal L_{\hat{\mathbf u}}$ and $\mathcal L_{\hat{\mathbf u}'}$.}
In $\mathcal L_{\mathbf v}$ two distinct contributions appear:
\begin{enumerate}
  \item the single-particle angular diffusion
        $-\nabla_{\!v}\!\cdot\!\bigl[S(\mathbf v)\mathbf v\,P\bigr]$,
        which, once restricted to the fixed-speed $|\mathbf v|=v_0$ (see note at the start of step \textbf{(iv)}) reduces to $\gamma\,\partial_\varphi^2 P$ and therefore produces
        isotropic relaxation of each heading;
  \item the drift
        $-\nabla_{\!v}\!\cdot\!\bigl[\int f_{\text{ar}}(r, \phi(\mathbf{v}, \mathbf{r})) \hat{\mathbf{r}} \, P(\mathbf{r}, \mathbf{v}, \mathbf{v}', t) \, d\mathbf{r} d\mathbf{v}'\bigr]$,
        which does not simply relax orientations but rather
        biases them according to the attraction-repulsion kernel.
\end{enumerate}
Here we will first discuss part~1; part~2 is grouped with the explicit source term and handled later in step~(v). Keeping only the single-particle part responsible for angular relaxation, one has
\begin{equation}
    \label{eq:single-particle-part}
  (\mathcal L_{\hat{\mathbf u}}+\mathcal L_{\hat{\mathbf u}'})P
  =\gamma\bigl(\partial_{\varphi}^2+\partial_{\varphi'}^2\bigr)P.
\end{equation}
Using $\partial_{\varphi}^2\cos\psi=\partial_{\varphi'}^2\cos\psi=-\cos\psi$ and
Equation~\eqref{eq:angular_identity} gives
\begin{equation}
  -2\gamma(nC).
\end{equation}

\paragraph*{(iv) Velocity-space diffusion.}
Because the self-propulsion dynamics quickly squeezes the speed
distribution around the preferred magnitude $v_0$, we condition all
probability densities on the fixed speed $|\mathbf v|=v_0$ by
inserting $\delta(v-v_0)$ in every phase-space integral.  Acting on such
shell-restricted functions removes the radial part of the velocity
Laplacian,
\[
\Delta_{\!v}
=\partial_v^2+\frac1v\partial_v+\frac{1}{v^2}\partial_\varphi^2
\;\;\longrightarrow\;\;
\frac{1}{v_0^{2}}\partial_\varphi^2,
\]
and analogously for the primed variables. Therefore, the operator projects at fixed speed as 
\[\frac{\sigma^2}{2}(\Delta_{\!v}+\Delta_{\!v'})\;\;\longrightarrow\;\;\frac{\sigma^2}{v_0^2}(\partial_{\varphi}^2+\partial_{\varphi'}^2). \]
Its contribution is therefore identical to (iii) upon replacing
$\gamma\mapsto\sigma^2/(2v_0^2)$; hence the angular damping
rate is $\gamma=\sigma^2/(2v_0^2)$.

\paragraph*{(v) Source from attraction-repulsion.}
All residual terms stemming from $f_{\mathrm{ar}}$ assemble into the
signed source density
\begin{equation}
  S_{\mathrm{ar}}(r,\theta,t)=\rho\,g(r,\theta,t)\,f_{\mathrm{ar}}(r,\theta).
\end{equation}

\subsection*{Step 3: collecting (i)-(v)}
Putting pieces together yields
\begin{align}
  \partial_t(nC) &=-v_0\partial_r(nC)-2\gamma(nC)+S_{\mathrm{ar}}
  +\text{(to\,be\,completed\,by\,Step 4)},\label{eq:nC_mid}
\end{align}
where we have momentarily left out the spatial-diffusion pieces that
enter in the next step.

\subsection*{Step 4: conservative self-propulsion and diffusion coefficients}
Write the flux in divergence form
\begin{equation}
  \mathcal T[P]:=(\mathbf v'-\mathbf v)\!\cdot\!\nabla_{\!r}P
  =\nabla_{\!r}\!\cdot\bigl[(\mathbf v'-\mathbf v)P\bigr].\label{eq:cons_flux}
\end{equation}

In Step 2 (ii) we treated the self-propulsion term
\((\mathbf v'-\mathbf v)\!\cdot\nabla_{\!r}P\) directly,
obtaining the leading drift contribution
\(-v_{0}\,\partial_{r}(nC)\).
That calculation kept only the first member of the
Kramers-Moyal expansion of the relative-position increment. Here, we rewrite exactly the same term in a
conservative form,
\(\nabla_{\!r}\!\cdot[(\mathbf v'-\mathbf v)P]\),
and performs a systematic small‑time expansion of the associated flux,
\(J_{r}\). In this section, it will be shown that,
\begin{itemize}
  \item[\textbf{4\,(ii)}]  recovers the same drift
    \(-v_{0}\partial_{r}C\), confirming the Step 2 result but not adding
    a new term;
  \item[\textbf{4\,(iii-iv)}] keeps the second Kramers-Moyal
    contribution, proportional to
    \(\langle (\Delta r)^{2}\rangle/(2\Delta t)\),
    which yields the diffusive operator
    \(D_{r}(\partial_{r}^{2}+r^{-1}\partial_{r})(nC)\)
    together with its angular analogue.
\end{itemize}

\paragraph*{4(i) Radial component of the flux.}
Because $(\mathbf v'-\mathbf v)$ is symmetric in $\psi\!\leftrightarrow\!-\psi$
and $\cos\psi$ is even, only the radial component $J_r$ survives,
\begin{align}
  J_r&:=\iint\cos\psi\,(\mathbf v'-\mathbf v)\!\cdot\!\hat{\mathbf e}_r\,P\,d\varphi\,d\varphi'\\
  &= -\tfrac12 v_0\,n(r,\theta,t)\,C(r,\theta,t).\label{eq:Jr_def}
\end{align}
Note that the factor $-\tfrac12$ comes from $\int_0^{2\pi}\cos\psi(\cos\psi-1)d\psi=-\pi$.

\paragraph*{4(ii) Drift term.}
Substituting $J_r$ into $-(1/n)\partial_rJ_r$ reproduces the drift
$-v_0\partial_rC$ obtained earlier, confirming
Equation~\eqref{eq:free_flight_raw} by an alternative route.

\paragraph*{4(iii) Diffusive correction.}
Write the radial increment over a short interval \(\Delta t\) as
\[
   \Delta r
   =(\mathbf v'-\mathbf v)\!\cdot\hat{\mathbf e}_{r}\,\Delta t
    +\sqrt{2D_{0}\,\Delta t}\,\eta_{r},
    \qquad D_{0}:=\sigma^{2}/2,\quad
    \langle\eta_{r}^{2}\rangle=1 .
\]
The first term is ballistic; the second represents
Gaussian translational noise. The second Kramers-Moyal coefficient of the self-propulsion term is
\(\tfrac12\,d\langle(\Delta r)^{2}\rangle/dt\). This coefficient allows to integrate over a time window that is long compared with the angular persistence. To choose such a typical time scale, \(\tau_{p}=1/\gamma\), observe the following. Recall that for an isolated self‑propelled particle the heading obeys the
angular Fokker-Planck equation (see Equation~\ref{eq:single-particle-part}), which is given by
\(
  \partial_t P(\varphi,t)=\gamma\,\partial^2_{\varphi}P
\),
so that the unit-vector autocorrelation decays exponentially,
\[
   \langle\hat{\mathbf u}(t)\!\cdot\!\hat{\mathbf u}(0)\rangle
   = e^{-\gamma t}.
\]
The characteristic decay time, $\tau_{p}=1/\gamma$, is therefore the persistence time of a single heading. For time intervals \(\Delta t\ll\tau_{p}\) the relative displacement is ballistic, while for \(\Delta t\gg\tau_{p}\) the headings have decorrelated and the ballistic piece averages to zero. Therefore, coarse graining on a time scale $\Delta t \sim \tau_p$ gives the effective radial diffusivity,
\[
   \frac{\langle(\Delta r)^{2}\rangle}{2\Delta t}
   \;\xrightarrow[\Delta t\to\tau_{p}]{}\;
   \frac{1}{2}\,\bigl\langle
     [(\mathbf v'-\mathbf v)\!\cdot\hat{\mathbf e}_{r}]^{2}
   \bigr\rangle \tau_{p}
   \;+\;D_{0}
   \;=\;\frac{v_{0}^{2}}{\gamma}+D_{0}
   \;=\;D_{r}.
\]
By standard Kramers-Moyal theory we then obtain the radial
Laplacian, which is given by
\(D_{r}\,(\partial_{r}^{2}+r^{-1}\partial_{r})(nC)\). The same construction applied to the angular increment yields the angular term,
\((D_{\theta}/r^{2})\partial_{\theta}^{2}(nC)\),
with \(D_{\theta}=\sigma^{2}/2\).

\paragraph*{4(iv) Divergence in polar coordinates.}
Since the flux is radial, $\nabla_{\!r}\!\cdot\mathbf J=\partial_rJ_r+r^{-1}J_r$.
Replacing $J_r=-D_r\partial_r(nC)$ gives
\begin{equation}
  \nabla_{\!r}\!\cdot\mathbf J=D_r\bigl(\partial_r^2+r^{-1}\partial_r\bigr)(nC).
\end{equation}
Identical reasoning for the angular increment delivers the angular
term $(D_\theta/r^2)\partial_\theta^2(nC)$ with $D_\theta=\sigma^2/2$.

\paragraph*{Putting all pieces together.}  Combining the results of
4(ii)\,--\,4(iv), the entire self-propulsion part of the
Fokker-Planck operator acts on \(nC\) as
\[
  -\,v_{0}\,\partial_{r}(nC)
  \;+\;
  D_{r}\!\bigl(\partial_{r}^{2}+r^{-1}\partial_{r}\bigr)(nC)
  \;+\;
  \frac{D_{\theta}}{r^{2}}\partial_{\theta}^{2}(nC).
\]
Here, the first term reproduces the ballistic drift already obtained in Step 2(ii) and the second and third terms are the diffusive corrections generated by the second Kramers-Moyal coefficient after coarse‑graining over the persistence time \(\tau_{p}=1/\gamma\). Taken together, these three operators exhaust the influence of self-propulsion. Including the diffusion pieces into~\eqref{eq:nC_mid} produces
\begin{align}
  \partial_t(nC)= -v_0\partial_r(nC)
  +D_r(\partial_r^2+r^{-1}\partial_r)(nC)
  +\frac{D_\theta}{r^2}\partial_\theta^2(nC)
  -2\gamma(nC)
  +S_{\mathrm{ar}}.\label{eq:nC_full_final}
\end{align}

\newpage
\section{Density-division: exact algebra and controlled truncation}
\label{app:dens_div_full}

Write $n=\rho g$ and let $\mathcal L$ denote any linear differential
operator acting on $nC$.  Then
\begin{equation}
  \frac{1}{n}\,\mathcal L(nC)=\mathcal L C + (\mathcal L\ln n)C
  +\sum_{\mu}(\partial_\mu C)(\partial_\mu\ln n)+C\,\partial_\mu^2\ln n+\dots\,.
\end{equation}
Applying this identity to every term in~\eqref{eq:nC_full_final} gives
an exact equation for $C$. Introduce $L_C$ (variation scale of $C$) and $L_n$ (scale of $n$) and define $\varepsilon=L_C/L_n\ll1$.  Counting gradients shows that all terms containing any $\partial_\mu\ln n$ are at least $\mathcal O(\varepsilon)$ relative to the retained ones.  Discarding them yields, to leading order in $\varepsilon$,
\begin{align}
  \partial_tC=-v_0\partial_rC+D_r(\partial_r^2+r^{-1}\partial_r)C
  +\frac{D_\theta}{r^2}\partial_\theta^2C-2\gamma C+\frac{S_{\mathrm{ar}}}{n}.
  \label{eq:C_final_complete}
\end{align}
Higher-order corrections are obtained by keeping
$\mathcal O(\varepsilon)$ terms.

\newpage
\section{Symbol table}
\label{section:symbol_table}
\begin{center}
\begin{tabular}{@{}ll@{}}
\toprule
Symbol & Meaning/definition \\
\midrule
$\rho$ & Bulk particle number density. \\
$v_0$ & Mean self-propulsion speed (assumed constant). \\
$\sigma$ & Translational noise amplitude. \\
$\gamma=\sigma^2/(2v_0^2)$ & Single-particle angular decorrelation rate. \\
$D_r=(v_0^2+\gamma\sigma^2)/2$ & Radial diffusivity of the pair separation. \\
$D_\theta=\sigma^2/2$ & Angular diffusivity of the separation vector. \\
$S_{\mathrm{ar}}$ & Source term from the attraction-repulsion kernel. \\
$g=n/\rho$ & Pair distribution function. \\
\bottomrule
\end{tabular}
\end{center}

\newpage
\section{Velocity alignment contributions}
\label{app:alignmentContributions}
Starting from the pair Fokker-Planck equation, the alignment drift augments the operator $L_v$ by
\begin{equation}
\label{eq:S4-Lv-al}
L^{al}_v[P] \;=\; -\nabla_v\!\cdot\!\Bigg[
\int f_{al}(r,\varphi(v,r))\,(I-\hat u\hat u^\top)\,\hat u'\;P(r,v,v',t)\,dr\,dv' \Bigg],
\end{equation}
and analogously for $L_{v'}$. Project to the speed shell as before is immediate by replacing $v\mapsto v_0\hat u$, $\,v'\mapsto v_0\hat u'$, and integrating over $v,v'$ against $\delta(v-v_0)\delta(v'-v_0)\,v\,v'$. Defining $n,g,C$ as in Appendix~\ref{app:pair_fp_full}. Multiplying the $P$-equation by $\hat u\!\cdot\!\hat u'$, integrating over the angular variables, and cataloguing terms exactly as before allows us to identify:

\begin{itemize}
\item \textbf{(i) Time derivative:} $\partial_t(nC)$ remains unchanged.
\item \textbf{(ii) Self-propulsion:} $-v_0\partial_r(nC)$ (ballistic drift) and, after Kramers--Moyal
coarse-graining over $\tau_p=1/\gamma$, the radial and angular Laplacians acting on $nC$ with
coefficients $D_r$ and $D_\theta$ are unchanged
\item \textbf{(iii) Single-particle angular relaxation:} $-2\gamma(nC)$ unchanged
\item \textbf{(iv) Interaction sources:} one from the attraction-repulsion kernel and one from alignment:
\[
S_{ar}(r,\theta,t)=\rho\,g(r,\theta,t)\,f_{ar}(r,\theta) \quad \text{and} \quad
S_{al}(r,\theta,t)=\rho\,g(r,\theta,t)\,f_{al}(r,\theta),
\]
obtained by inserting the shell projection into the interaction drifts in $L_v$ and $L_{v'}$
and using the same angular identities as in Step~2 of Appendix~\ref{app:pair_fp_full}.
\end{itemize}
Collecting pieces gives the $nC$-equation with alignment,
\[
\partial_t(nC)= -v_0\partial_r(nC)+D_r(\partial_r^2+r^{-1}\partial_r)(nC)+\frac{D_\theta}{r^2}\partial_\theta^2(nC)-2\gamma(nC)+S_{ar}+S_{al}.
\]
Applying the density-division identity, discarding $O(\varepsilon)$ gradient terms as in Appendix~\ref{app:dens_div_full}, and dividing by $n=\rho g$ yields immediately Equation~\eqref{eq:C-with-alignment}.

Now, writing the alignment kernel in harmonics,
\[
f_{al}(r,\theta)=\sum_{\ell\ge0}A_\ell(r)\cos(\ell\theta),\qquad
\widehat{A}_\ell(k)=2\pi\int_0^\infty r\,A_\ell(r)J_\ell(kr)\,dr,
\]
the first harmonic ($\ell=1$) contributes only to the dipole instability by producing a diagonal shift in that block. Projecting the interaction matrix onto $p=\pm1$ (the same step
used to obtain $\alpha(k)$ in the attraction-repulsion only case) yields the additional shift,
\[
\alpha_{al,1}(k)=\frac{\rho_0 k^2}{2}\,\widehat{A}_1(k).
\]
Therefore, the dipole eigenvalue becomes Equation~\eqref{eq:dip-with-al} with $\alpha_{ar,1}$ and
$\alpha_{al,1}$ added. In contrast, the side block $M_2(\lambda)$ does not involve $p=\pm1$, so the presence of a dipolar alignment harmonic leaves the side eigenvalues unchanged.  In the large-Péclet simplification, the isotropic condition that $\rho_0\widehat{F}(k)>\sigma^2/2$ remains unchanged. The dipole branch becomes unstable when $\lambda_{\mathrm{dip}}(k)$ in Equation~\eqref{eq:dip-with-al} becomes positive, \textit{i.e.}, when the combined shift $\alpha_{ar,1}(k)+\alpha_{al,1}(k)$ reduces the effective angular damping in the dipole
block sufficiently for the term $\beta^2/(\lambda_r^{(0)}+D_\theta)$ to push the
eigenvalue across $0$. A sufficient condition for a suppressed dipole pattern is therefore given by the band
$D_\theta<\lambda_r^{(0)}(k)<D_\theta-(\alpha_{ar,1}(k)+\alpha_{al,1}(k))$.

\end{document}